%% file: Complete_state_tomography_of_a_quantum_dot_spin_23.tex
\documentclass[twoside,english,reprint, aps, prx, superscriptaddress, amsfonts, amssymb, amsmath, a4paper]{revtex4-1}
\usepackage{lmodern}

\usepackage[LGR,T1]{fontenc}
\usepackage[latin9]{inputenc}
\usepackage{color}
\usepackage{babel}
\usepackage{textcomp}
\usepackage{amsmath}
\usepackage{amssymb}
\usepackage{graphicx}
\usepackage[unicode=true,
 bookmarks=false,
 breaklinks=true,pdfborder={0 0 0},pdfborderstyle={},backref=false,colorlinks=true]
 {hyperref}
\hypersetup{pdftitle={Relaxation Dynamics of Spin Qubits in Quantum Dots},
 pdfauthor={Dan Cogan}}

\makeatletter

\DeclareRobustCommand{\greektext}{%
  \fontencoding{LGR}\selectfont\def\encodingdefault{LGR}}
\DeclareRobustCommand{\textgreek}[1]{\leavevmode{\greektext #1}}
\ProvideTextCommand{\~}{LGR}[1]{\char126#1}

\newcommand{\lyxmathsym}[1]{\ifmmode\begingroup\def\b@ld{bold}
  \text{\ifx\math@version\b@ld\bfseries\fi#1}\endgroup\else#1\fi}

\makeatother

\begin{document}
\title{Complete state tomography of a quantum dot confined spin qubit.}
\author{Dan Cogan}
\affiliation{The Physics Department and the Solid State Institute, Technion\textendash Israel
Institute of Technology, 3200003 Haifa, Israel}
\author{Giora Peniakov}
\affiliation{The Physics Department and the Solid State Institute, Technion\textendash Israel
Institute of Technology, 3200003 Haifa, Israel}
\author{Zu-En Su}
\affiliation{The Physics Department and the Solid State Institute, Technion\textendash Israel
Institute of Technology, 3200003 Haifa, Israel}
\author{David Gershoni}
\email{dg@physics.technion.ac.il}

\affiliation{The Physics Department and the Solid State Institute, Technion\textendash Israel
Institute of Technology, 3200003 Haifa, Israel}
\begin{abstract}
\textcolor{black}{Semiconductor quantum dots are probably the preferred
choice for interfacing anchored, matter spin qubits and flying photonic
qubits. While full tomography of a flying qubit or light polarization
is in general straightforward, matter spin tomography is a challenging
and resource-consuming task. Here we present a novel all-optical method
for conducting full tomography of quantum-dot-confined spins. Our
method is applicable for electronic spin configurations such as the
conduction-band electron, the valence-band hole, and for electron-hole
pairs such as the bright and the dark exciton. We excite the spin
qubit using short resonantly tuned, polarized optical pulse, which
coherently converts the qubit to an excited qubit that decays by emitting
a polarized single-photon. We perform the tomography by using two
different orthogonal, linearly polarized excitations, followed by
time-resolved measurements of the degree of circular polarization
of the emitted light from the decaying excited qubit. We demonstrate
our method on the dark exciton spin state with fidelity of 0.94, mainly
limited by the accuracy of our polarization analyzers.}
\end{abstract}
\maketitle
\global\long\def\ket#1{\left|#1\right\rangle }%
\global\long\def\im{\operatorname{Im}}%
\global\long\def\bra#1{\left\langle #1\right|}%

\section{Introduction}

\begin{figure}
\includegraphics[width=1\columnwidth]{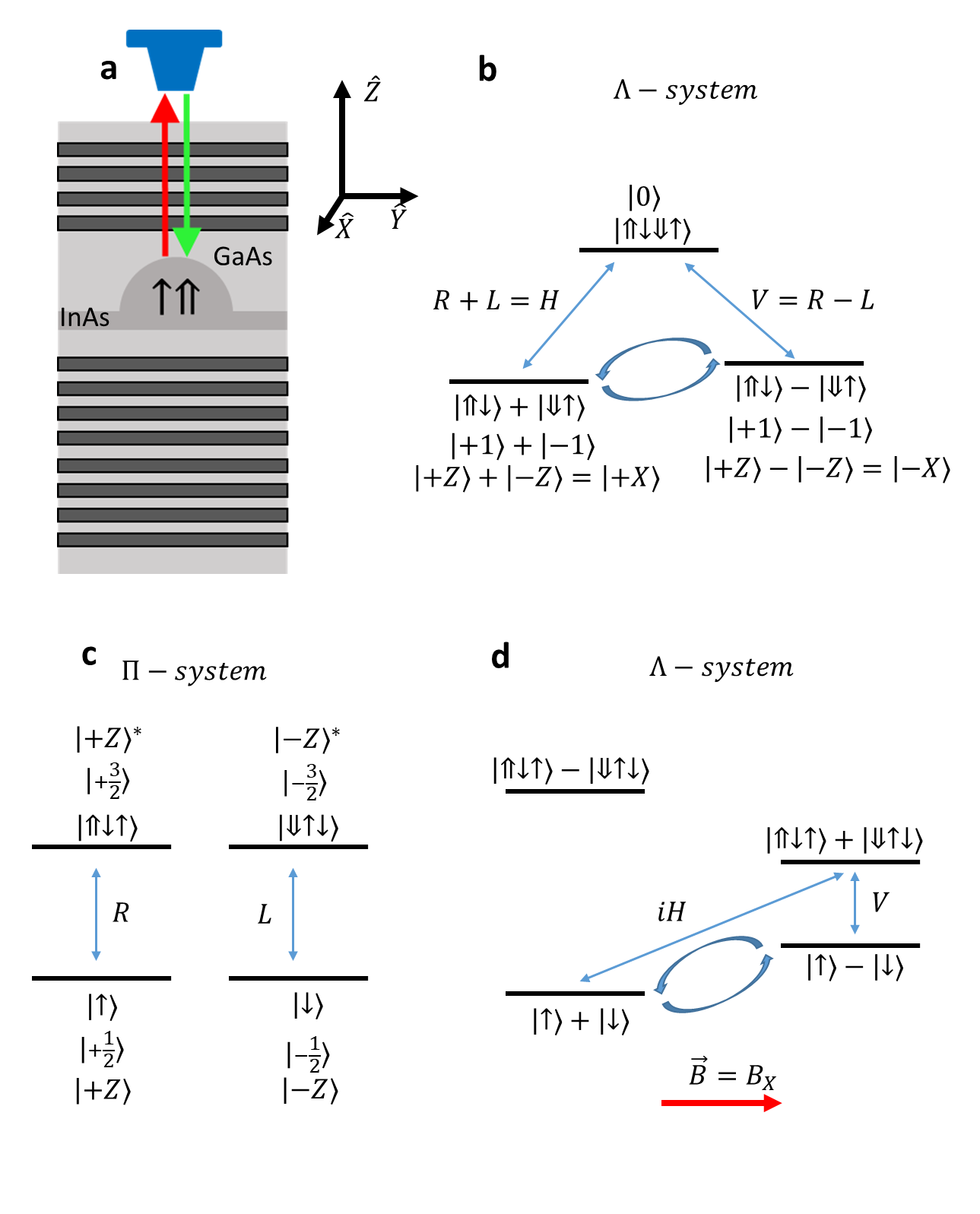}

\caption{\label{fig:Schematical-description-1-1}a) Schematic description of
the studied single QD device. The QD is located in the focal point
of a microscope objective, which focuses the pulsed laser beam (represented
by a green arrow) and collimates the emitted PL (red arrow). Here
$\hat{z}$ defines the growth direction of the QD. b) schematic description
of the confined BE - biexciton as an optical $\Lambda-$system and
c) the confined electron -trion as an optical $\Pi-$system. d) Transforming
an optical $\Pi-$system into a $\text{\textgreek{L}}-$system using
an external magnetic field. Here $\uparrow$ ($\Downarrow$) represents
spin-up electron (spin-down hole) and $R$ and $L$ represent right
and left hand circularly polarized optical transition. Numbers indicate
the total angular momentum of the electronic state. Here |$\protect\ket{+Z}$
($\protect\ket{-Z}^{*}$) represent spin up ground (spin down excited
) state along the $\hat{z}$ axis, and $\protect\ket{+X}$ ($\protect\ket{-X}^{*}$)
represent coherent superpositions as defined in the figure.}
\end{figure}

Qubits are the building blocks of quantum technologies \citep{Loss1998,DiVincenzo_2000}.
Qubits can be realized in many physical two-level systems, which maintain
their coherence for a much longer time than the time required to initialize
their coherent states, read, or subject the state to logical gates
\citep{Kane1998,Bloch2008,Clarke2008,Hanson2008}. The vision of quantum
technologies is frequently described in terms of anchored qubits,
located on different nodes in space where they serve as quantum information
processors or quantum memories, and flying qubits that propagate long
distances, connecting the various nodes\citep{Kimble2008}. Photons
are natural flying qubits since they can travel long distances without
dephasing, while their polarization state carries the quantum information
\citep{Imamog_lu_1999,Sangouard2011}. The spins of single electrons,
nuclei, atoms or molecules are examples for anchored qubits. In some
instances, they can be isolated, thereby maintaining spin coherence
for very long times \citep{Johnson2005,Bloch2008,Blatt2008,Saeedi2013,Cogan2018}.
Semiconductor quantum dots are an excellent interface between anchored
spin qubits and flying photonic qubits due to their ability to both
isolate electronic spin qubits and to enhance their interaction with
the photons light field. Quantum photonic devices based on semiconductor
quantum dots, such as described in Fig. \ref{fig:Schematical-description-1-1}a,
are almost ideally suited for building highly-performant, optically-active
quantum nodes.

\textcolor{black}{Determining the qubit state generally requires projections
on various basis states in a process called tomography. Full tomography
of the polarization state of flying photonic qubits is quite straightforwardly
done using state of the art polarizers and retarders \citep{Winik2017,Wang2018a}.
Researchers have developed methods for measuring matter qubits' spin
in various systems \citep{Kosaka2009,Morello2010,Neumann2010,Pla2013,Nakajima2017}.
Full tomography of confined spin qubits in semiconductor QDs remains
however, challenging and demanding \citep{Kloeffel2013,Gywat2009}.
These difficulties are reflected in the various methods, developed
in recent years, which require application of relatively strong magnetic
fields. The fields affect the qubits and alter their energy level
structure \citep{Greve2013,Gao2012,Greve2012}, rendering the spin
tomography inaccurate.}

Relevant to this work is the all-optical method developed by Benny
et al. \citep{Benny2011}, demonstrating full tomography of a QD confined
bright exciton (BE - an electron-hole pair) spin. The bright exciton
is a spin integer quasi-particle, with a total angular projection
of either +1 or -1 on the QD symmetry axis. This nondegenerate spin
qubit forms an optical $\Lambda-$system with the spin-zero biexciton
state, which results from resonant optical excitation of the BE, as
schematically described in Fig. \ref{fig:Schematical-description-1-1}b.
In this optical $\Lambda-$system, each of the BE qubit's two states
is optically connected to the biexciton single state. A nondegenerate
qubit, which is part of an optical $\Lambda-$system facilitates a
relatively easy way for optical spin tomography, resulting from a
one-to-one correspondence between the polarization of the spin and
the photon, inducing the optical transition. A nondegenerate spin
qubit naturally precesses in time, and therefore, time-resolved polarization-sensitive
spectroscopy can be used for full tomography of the precessing spin
qubit \citep{Benny2011}.

The situation is different for a QD confined single-charge carrier
such as the conduction-band electron or the valence-band hole. A charge
carrier has a half-integer spin, and as such in the absence of an
external magnetic field, the qubit that it forms is Kramers' degenerate.
Similarly, an optically excited charge carrier (positive or negative
trion), is also a Kramers' degenerate half-integer spin state. Therefore,
the charge and the excited charge are both spin qubits that form an
optical $\Pi-$system rather than a $\Lambda-$system as can be seen
in Fig. \ref{fig:Schematical-description-1-1}c. The conduction-band
electron \citep{Loss1998,DiVincenzo_2000,Kroutvar2004}, the valence-band
hole \citep{Brunner2009,De_Greve_2011}, and the dark exciton (DE-
optically inactive electron-hole pair)\citep{Poem2010,Schwartz2015,Schwartz2015a}
are long-lived ground electronic spin qubits in a QD system. The first
two are Kramers' degenerate, but the third one is not. All three qubits
form a natural $\Pi-$system with circularly polarized optical selection
rules to their excited qubits \citep{Cogan2018}. Thus, tomography
as in the BE case \citep{Benny2011} is impossible.

One way to circumvent this problem as described in Fig. \ref{fig:Schematical-description-1-1}d
is to apply a strong magnetic field, which lifts the Kramers' degeneracy
of the optically excited spin state, thereby facilitating an optical
$\Lambda-$like system between the carrier's spin states and one of
the optically excited states \citep{Greve2013,Togan2010,Gao2012}.
However, the strong magnetic field also lifts the degeneracy of the
ground level qubit, thus creating a spectral \textquotedbl which
path\textquotedbl{} information for the emitted photons. Therefore,
the use of this artificially-created $\Lambda-$system for spin-photon
entanglement \citep{Greve2012,Gao2012}, requires erasing the spectral
or the polarization information encoded in the emitted photon, thereby
limiting the method applicability and scalability.

Here, we develop and demonstrate a novel, all-optical spin tomography
method, which uses the natural $\Pi-$system that the spin qubit is
a part of, for the optical tomography. The method is general, accurate,
and can be applied for state-tomography of multi-qubits, and entangled
spin-photon states \citep{Lindner2009,Economou2010,Schwartz2016}.
For Kramers\textquoteright{} degenerate spin qubits, however, a minimal
external magnetic field in Voigt configuration is still required for
lifting the degeneracy and induce spin precession. We perform the
tomography in the following way: first, we excite the spin qubit under
study, which we call the ground level qubit, to an optically active
excited qubit state, which we call the excited qubit. This coherent
and deterministic conversion is done using a short resonantly-tuned
linearly-polarized optical $\pi-$pulse, such that there is a one-to-one
correspondence between the state of the ground level qubit and that
of the excited qubit. The excited qubit then radiatively recombines
while its spin state evolves during the recombination in a frequency
given by the energy difference between the excited qubit eigenstates.
The $\Pi-$system optical selection rules are such that by using time-resolved
circular polarization-sensitive PL spectroscopy one can trace back
the state of the excited qubit at the moment of conversion, and in
turn the state of the ground level qubit, prior to the conversion
pulse. Full tomography is obtained using two conversions on two different
linear polarization bases.

Unlike the single charge carriers, the DE is an integer spin qubit
\citep{Poem2010,Schwartz2015}, its eigenstates are not degenerate,
and thus even in the absence of an external magnetic field, the DE
qubit naturally precesses. In addition, the DE spin state can be written-up
as any desired coherent superposition of its eigenstates by a single,
picosecond long optical pulse \citep{Schwartz2015a}. Therefore, we
demonstrate the tomography on the DE spin qubit, though the method
applies to single-charge carriers as well.

\section{Experiment}

Typical size of an epitaxially grown semiconductor QD is tens of nanometers
in diameter, and a few nanometers in height, forming a 3D potential
trap that confines electrons, holes, and electron-hole pairs (excitons).
Figure \ref{fig:Schematical-description-1-1}a illustrates the QD
device used in this work. The InAs layer of the QDs is embedded in
a GaAs optical microcavity formed by 2 Bragg reflecting mirrors of
Al- GaAs/GaAs. The role of the microcavity is to increase the harvesting
efficiency of the light emitted from the QD \citep{Ramon2006}. A
microscope objective with a numerical aperture of 0.8 is used to both
focus the laser pulses on a single QD, and to collect the single photons
emitted. A QD confined dark exciton (DE) is composed of a conduction-band
electron and a valence-band hole pair, with parallel spins \citep{Poem2010,Schwartz2015}.
It is called dark since the two carriers cannot recombine radiatively
due to their electronic spin mismatch. This fact is also reflected
in the DE total angular momentum projection of $\pm2$ which can not
be carried out by a single photon, possessing an angular momentum
of $\pm1,$ only. Consequently, the QD confined DE forms a spin qubit
which has orders of magnitude longer lifetime than the BE \citep{Schwartz2015}.

\begin{figure}
\includegraphics[width=1\columnwidth]{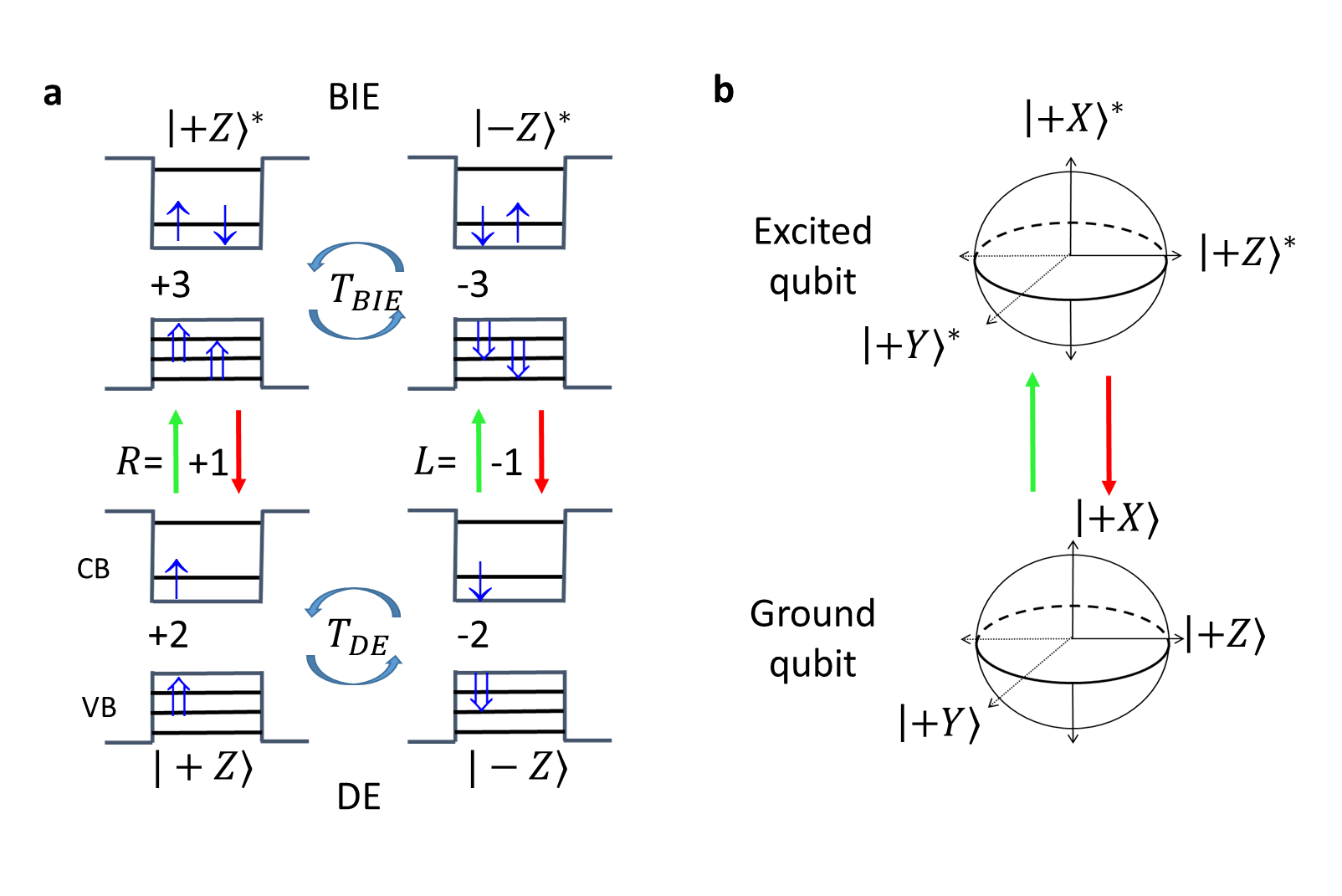}

\caption{\label{fig:Schematical-description-1} a) Energy levels, spin wavefunctions,
and polarization selection rules for resonant optical transitions
for the QD confined dark exciton (DE) and the optical P-system that
it forms with its optically excited biexciton (BIE). (b) Bloch sphere
representation of the confined ground spin qubit and the excited spin
qubit. Here $\protect\ket{+Z}$ ($\protect\ket{-Z}^{*}$) represent
spin up ground (spin down excited ) state along the $\hat{z}$ axis,
and $\protect\ket{+X}$ ($\protect\ket{-X}^{*}$) and $\protect\ket{+Y}$
($\protect\ket{-Y}^{*}$) represent coherent superpositions.}
\end{figure}

The DE can be optically excited by absorbing a photon, thereby photogenerating
an additional electron-hole pair in the QD. Resonant excitation of
the DE as schematically described in Fig. \ref{fig:Schematical-description-1}a
will result in an excited biexcitonic spin qubit (BIE). The BIE is
a four carrier state comprised of two anti-parallel electron spins
forming a singlet in the lowest conduction band level and two heavy-holes
with parallel spins, forming a triplet, where one hole is in the valence
bands' ground energy level, and the other one is in the first excited
energy level. The total angular momentum of the BIE is $\pm3,$ given
by the total spin projection of the two unpaired holes. Figure \ref{fig:Schematical-description-1}b
illustrates the Bloch sphere representation of the confined ground
spin qubit and the excited spin state. $\ket{+Z}$ ($\ket{-Z}^{*}$)
represent spin up ground (spin down excited) state along the $\hat{z}$
axis, defined by the growth direction of the QD, and $\ket{+X}$ ($\ket{-X}^{*}$)
and $\ket{+Y}$ ($\ket{-Y}^{*}$) represent coherent superpositions
as illustrated in the figure. After photogeneration, in about 0.5
ns, the excited spin decays radiatively by emitting a photon. One
note in Fig. \ref{fig:Schematical-description-1-1}c, and Fig. \ref{fig:Schematical-description-1}b
that the ground and excited qubit form an optical -system, in which
right (left) hand circularly polarized optical transitions connect
the $\ket{+Z}(\ket{-Z})$ states of the ground and excited qubits.

We use a linearly polarized resonant optical $\pi-$pulse to convert
the ground qubit coherently and deterministically to the excited qubit,
with a one-to-one correspondence between the two states. The excited
qubit then radiatively decays while its spin state evolves during
the relaxation. The selection rules of the optical $\Pi-$system are
such that a spin-up state of the excited qubit ($\ket Z^{*}$) results
in the emission of a right hand circularly polarized photon while
a spin-down state ($\ket{-Z}^{*}$) results in the emission of a left
hand circularly polarized photon. Thus, by using time-resolved circularly
polarized sensitive PL spectroscopy one can trace back the state of
the excited qubit at the moment of conversion, and in turn the state
of the ground-level qubit, prior to the conversion.

\begin{figure}
\includegraphics[width=1\columnwidth]{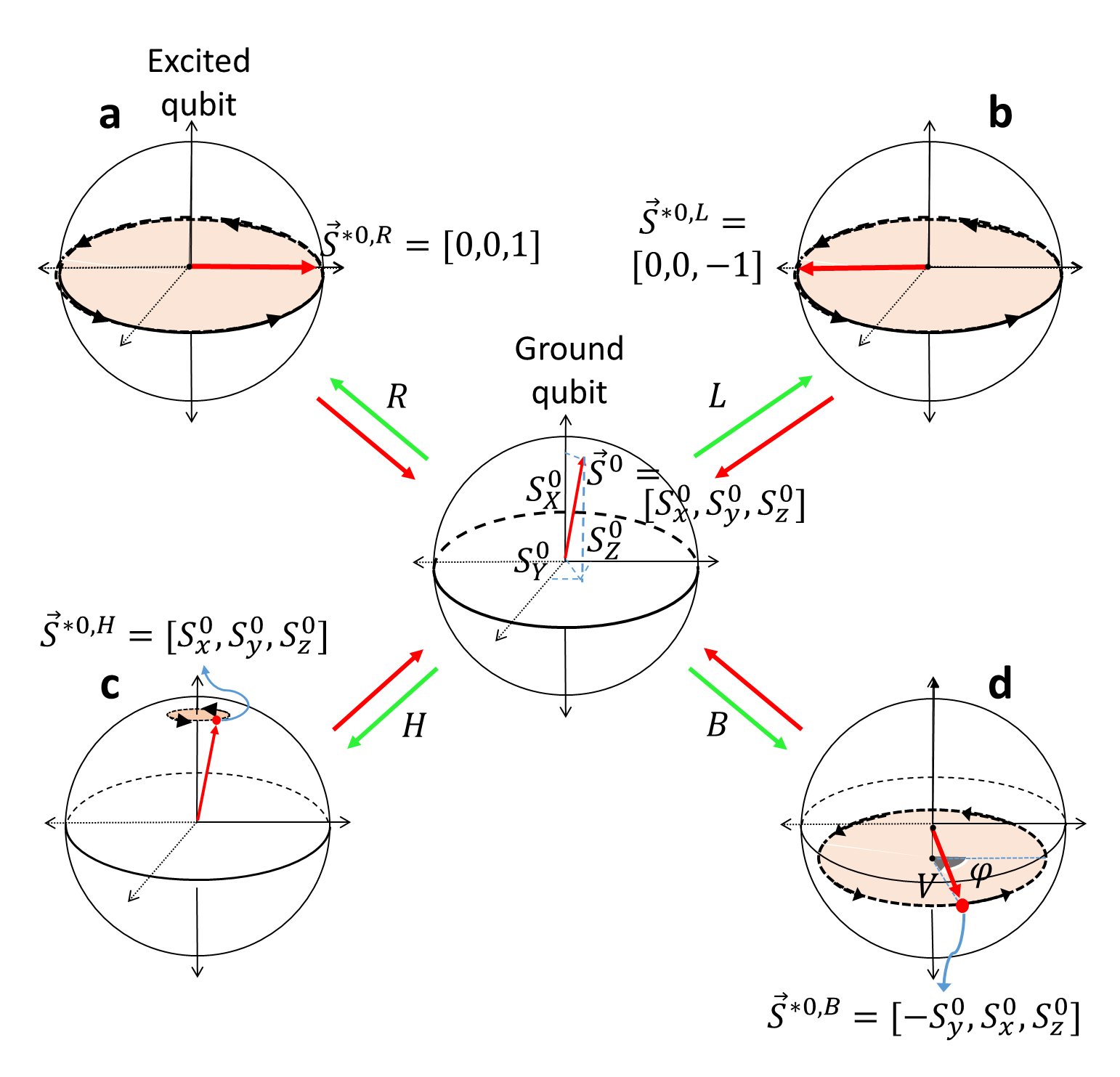}

\caption{\label{fig: Diagram of process}Schematic description of the spin-state
tomography. The red arrow on the central Bloch sphere represents the
confined ground-level spin-state, described by the vector $\vec{S^{0}}=[S_{X}^{0},S_{Y}^{0},S_{Z}^{0}]$.
A polarized optical $\pi-$pulse (green arrow) converts the ground
spin qubit into the excited qubit. The excited qubit evolves in time
while it radiatively recombines and emits a single photon (red arrow).
Figures (a)-(d) describe four different polarizations of the converting
pulse: (a) R-conversion, (b) L-conversion, (c) H-conversion, and (d)
B-conversion. In each case, the red arrow on the excited-qubit Bloch
sphere describes the excited-qubit state at the moment of conversion
and the direction of the state's precession. The amplitude $V^{0}$
and the time-dependent phase $\varphi(t)$ characterize this precession.
While R (L) circularly polarized excitation initializes the excited
qubit in its $\protect\ket{+Z}$ $(\protect\ket{-Z})$ spin-state,
H (B) polarized excitation maintains the phase information of the
ground qubit and converts it to $\vec{S^{*0,H}}=[S_{X}^{0},S_{Y}^{0},S_{Z}^{0}],$(
$\vec{S^{*0,B}}=[-S_{Y}^{0},S_{X}^{0},S_{Z}^{0}])$.}
\end{figure}

Fig. \ref{fig: Diagram of process} describes the procedure for full
state tomography. The central Bloch sphere describes a general ground-level-qubit
state as a coherent superposition of the two $\ket{\pm Z}$ states.
Likewise, the outer Bloch spheres describe the states of the excited
qubit as a superposition of the $\ket{\pm Z^{*}}$ states, following
the optical pulse conversions, for four different pulse polarizations.

\begin{figure}
\includegraphics[width=1\columnwidth]{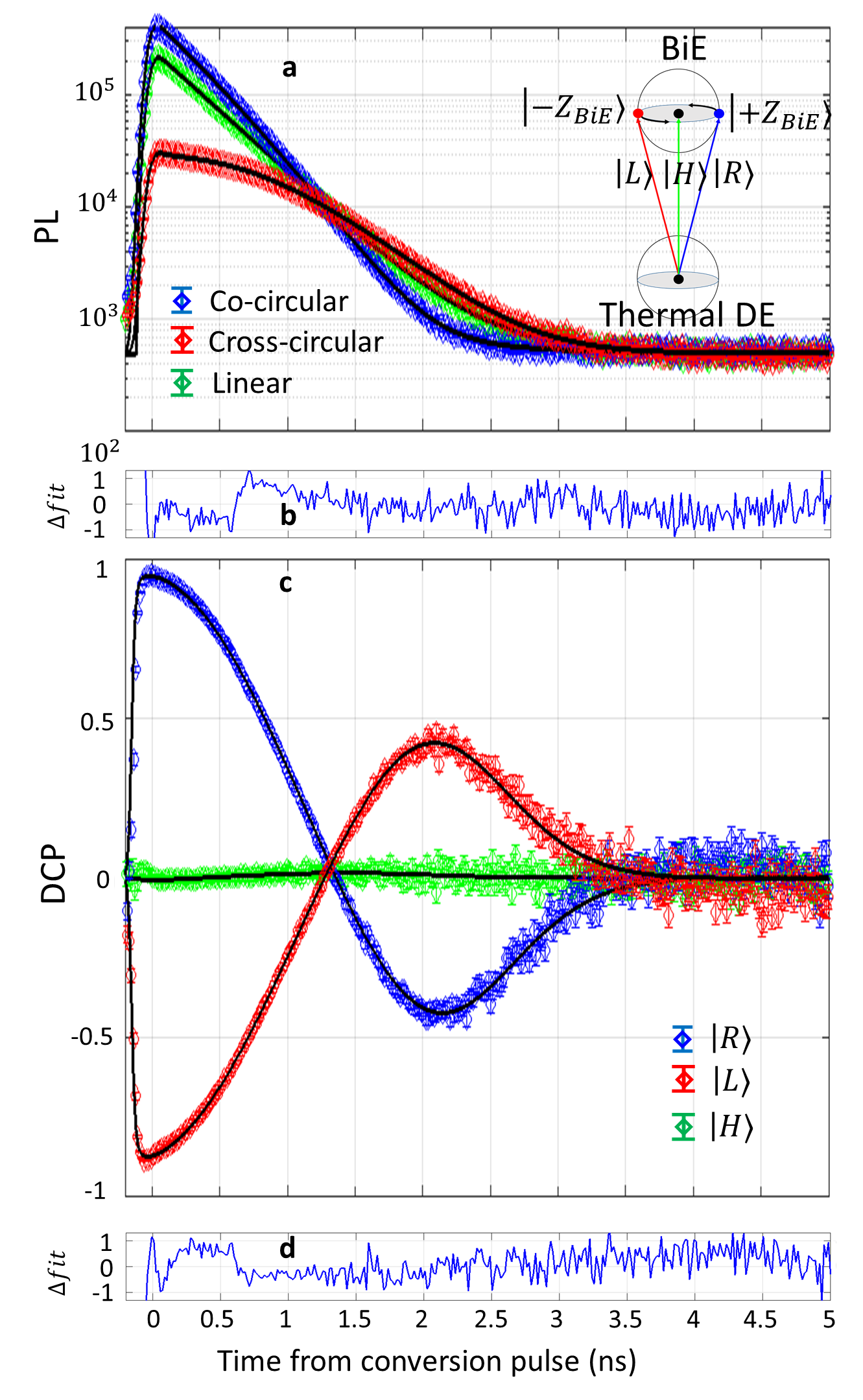}

\caption{\label{fig:characterization}Experimental characterization of the
BIE (excited DE) spin qubit. a) Polarization-sensitive time-resolved
PL measurement of the recombining BIE spectral line. Blue (red) diamond-marks
represent PL, detected in a co- (cross- ) circular-polarization to
the polarization of the excitation and green diamond marks represent
circularly polarized detection after linearly-polarized excitation.
The color-matched solid lines represent the best model fits to the
data using Eq. \ref{eq:PL R detection} and Eq. \ref{eq:PL L detection},
with three fitting parameters: $T_{BIE}=5.70\pm0.05ns$, $T_{2}^{*}=5.75\pm0.05ns$,
and $\tau_{R}=0.39\pm0.01ns$. (b) The difference between the fitted
model and the measured co-circular polarization data normalized by
the experimental uncertainty. (c) Time-resolved degree of circular
polarization (DCP) of the PL for R, L, and H polarized excitations.
(d) The difference between the fitted model and measured DCP for R
excitation, normalized by the experimental uncertainty.}
\end{figure}

\begin{figure*}
\begin{centering}
\includegraphics[width=1\textwidth]{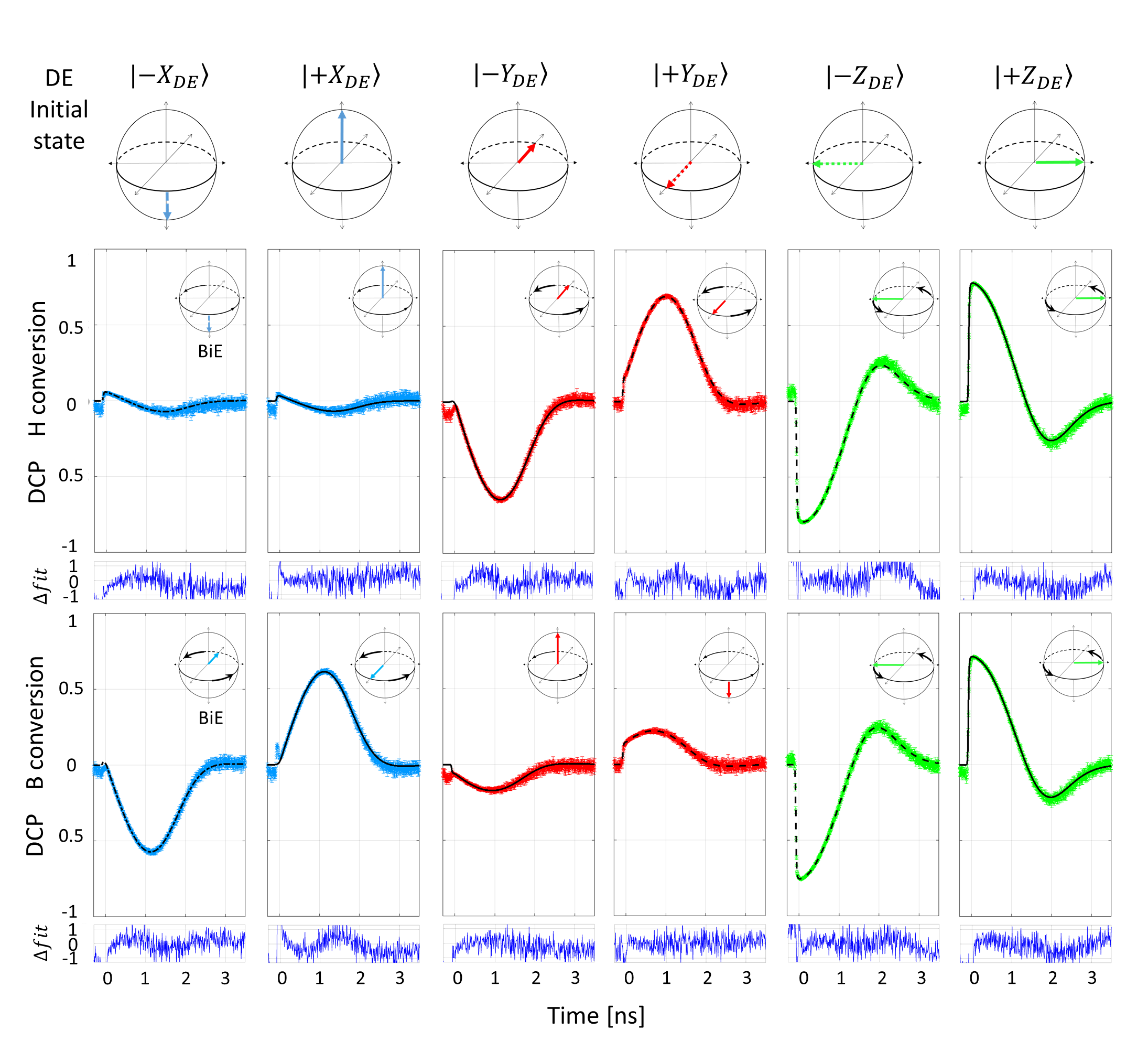}
\par\end{centering}
\caption{\label{fig:Tomographic measurements} Full state tomography of the
confined DE spin state for six different DE initializations, in 3
orthogonal bases. Blue, red and green arrows describe $\protect\ket{\pm X_{DE}}$
, $\protect\ket{\pm Y_{DE}}$, and $\protect\ket{\pm Z_{DE}}$ initializations
on the top DE-Bloch-spheres. Below each Initialization\textquoteright s
Bloch sphere, color-matched solid lines describe the measured time-resolved
$DCP(t)$ followed H and B conversions. Overlaid solid black lines
represent the best-fitted model using Eq. \ref{eq:Tomography} to
the measured data. The time-resolved differences between the fitted
model and the measured DCP, normalized by the experimental uncertainty,
are presented below each curve. The Bloch spheres in the insets describe
the initial BIE spin state after the conversion and its temporal precession
while it radiatively recombines.}
\end{figure*}

\begin{figure}
\includegraphics[width=1\columnwidth]{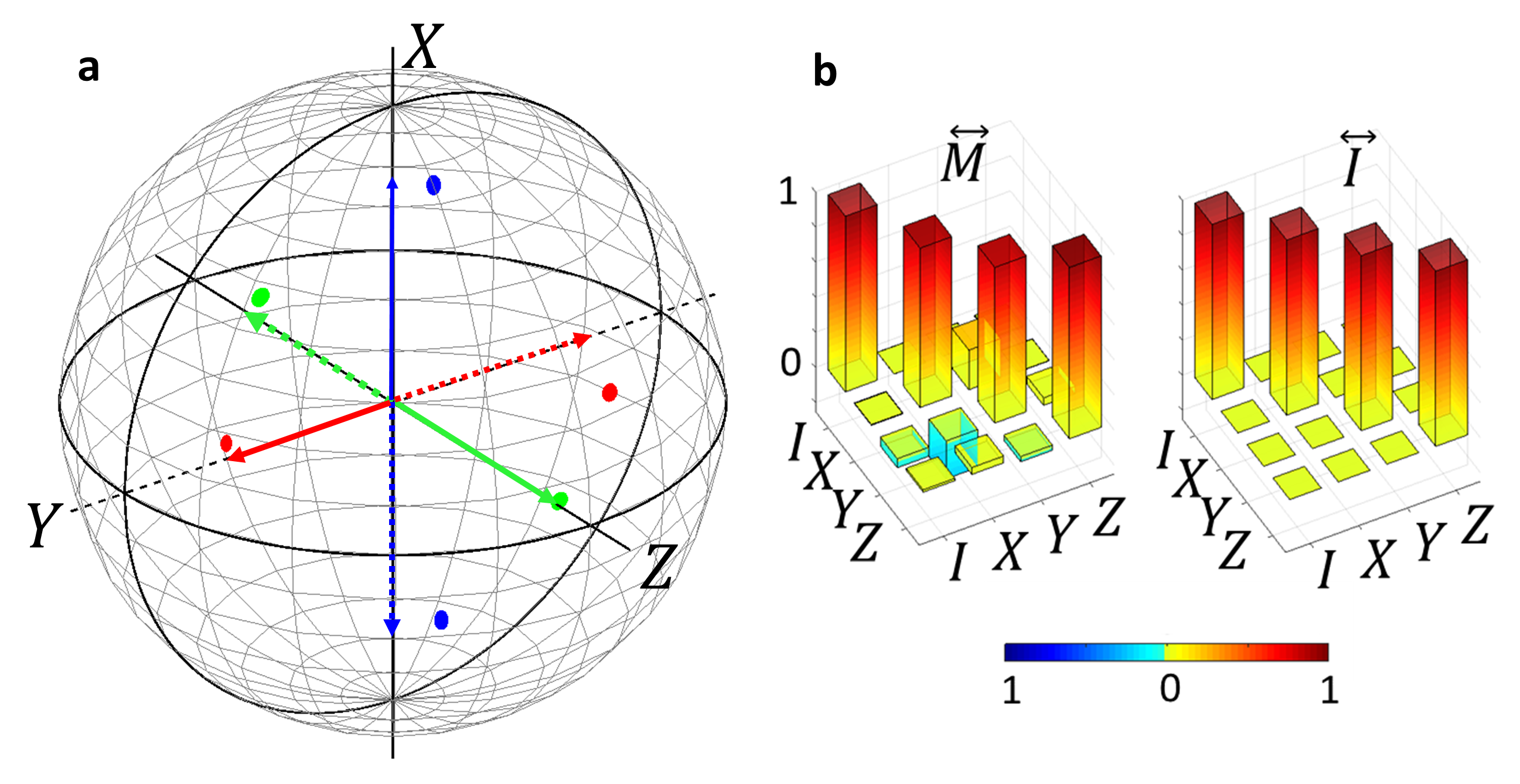}

\caption{\label{fig:Density matrices} a) DE Bloch sphere, representing the
initialized and measured states. Blue, red and green arrows represent
$\protect\ket{\pm X_{DE}}$, $\protect\ket{\pm Y_{DE}}$ and $\protect\ket{\pm Z_{DE}}$
DE initializations. The polarization degree is only 0.82 due to the
finite efficiency of the optical depletion. The color-matched spots
represent the tomographically measured states for each initialization.
The spots volume represents one standard deviation of the measurement
uncertainty. (b) Matrix representation ($\protect\overleftrightarrow{M}$)
of the state tomography measurement map. $\protect\overleftrightarrow{M}$
is a $4\times4$ positive-definite and trace-preserving map that maps
an initialized state (corrected for the 0.82 initialization) into
the measured one. The fidelity of the measured physical map, to the
identity map (also shown in b), is $0.94\pm0.02$.}
\end{figure}

In Fig. \ref{fig: Diagram of process}, as in Fig. \ref{fig:Schematical-description-1},
the green arrow describes the $\pi-$pulse which converts the ground-level-qubit
to the excited-qubit, while the red arrow describes single-photon
emission resulting from the excited-qubit recombination. The ground
spin state in Fig. \ref{fig: Diagram of process} is defined on the
central Bloch sphere by the vector: 
\[
\overrightarrow{S}^{0}=[S_{X}^{0},S_{Y}^{0},S_{Z}^{0}].
\]

$\overrightarrow{S}^{0}$ is the spin-state that one measures by state
tomography. This state is converted to the excited qubit by applying
a polarized 12-ps long \textminus pulse, energetically tuned to the
ground-level \textendash{} excited-level optical transition. Since
in this $\Pi-$system $\ket R$ photons connect only between $\ket{+Z}$
to $\ket{+Z^{*}}$ states, while $\ket L$ photons connect only between
$\ket{-Z}$ to $\ket{-Z^{*}}$ states, it follows that R- (L-) circularly
polarized conversion pulse, successfully applied to the ground-level-qubit
results in an excited qubit initial state which is given by $\vec{S^{*}}^{0,R(L)}=[0,0,\pm1]$
as shown in Fig. \ref{fig: Diagram of process}a (b). These two circularly
polarized conversion pulses always initialize the excited\textendash qubit
to the same states from any initial ground-level-qubit state. However,
when the conversion pulses are linearly polarized, the situation is
different. $\ket H=1/\sqrt{2}(\ket R+\ket L)$ polarized pulse converts
the ground-spin state to the same excited-spin state and $\ket B=1/\sqrt{2}\exp(-i\pi/4)(\ket R+i\ket L)$
linearly polarized pulse also rotates the excited-spin phase by $90\lyxmathsym{\textdegree}$
around the z-axis. Thus, after H conversion, the initial state of
the excited-qubit is given by: 
\begin{equation}
\vec{S^{*}}^{0,H}=[S_{X}^{0},S_{Y}^{0},S_{Z}^{0}],\label{eq:BiE-spin-H}
\end{equation}
as shown in Fig. \ref{fig: Diagram of process}c and after B conversion,
the initial state is: 
\begin{equation}
\vec{S^{*}}^{0,B}=[-S_{Y}^{0},S_{X}^{0},S_{Z}^{0}],\label{eq:BiE-spin-B}
\end{equation}
as shown in Fig. \ref{fig: Diagram of process}d.

Following the conversion the initial excited-qubit state evolves in
time by precessing around the eigenstates-axis ($X$) with a time
period $T_{excited}$. The precessing state projection on the Z-axis
of the excited-qubit Bloch sphere is therefore given by: 
\begin{equation}
S_{Z}^{*}(t)=V^{0}\cos(-\frac{2\pi t}{T_{excited}}+\varphi^{0}),\label{eq:BiE spin}
\end{equation}
where $V^{0}$ and $\varphi^{0}$ are defined as the amplitute and
phase. Here, for R conversion

\begin{equation}
V^{0}=V_{R}^{0}=1;\,\,\varphi^{0}=\varphi_{R}^{0}=0\label{eq:Hconversion-1}
\end{equation}

and for L conversion

\begin{equation}
V^{0}=V_{L}^{0}=1;\,\,\varphi^{0}=\varphi_{L}^{0}=\pi\label{eq:Hconversion-1-1}
\end{equation}

~~

while for H conversion

\begin{equation}
V^{0}=V_{H}^{0}=\sqrt{(S_{Y}^{0})^{2}+(S_{Z}^{0})^{2}};\,\,\,\varphi^{0}=\varphi_{H}^{0}=\arctan(S_{Y}^{0}/S_{Z}^{0})\label{eq:Hconversion}
\end{equation}

and for B conversion

\begin{equation}
V^{0}=V_{B}^{0}=\sqrt{(S_{X}^{0})^{2}+(S_{Z}^{0})^{2}};\,\,\,\,\varphi^{0}=\varphi_{B}^{0}=\arctan(S_{X}^{0}/S_{Z}^{0}).\label{eq:Bconversion}
\end{equation}

In addition to the natural coherent precession, the excited-qubit
state undergoes decoherence \citep{Bechtold2015,Cogan2018}, which
results in decay of the initial precession amplitude: 
\begin{equation}
V_{p}(t)=V_{P}^{0}\exp(-t^{2}/T_{2}^{*2}),\label{eq:Decoherence}
\end{equation}
where P=R, L, H or B and $T_{2}^{*}$ is the dephasing-time of the
excited-qubit. In addition, the excited-qubit decays radiatively while
its spin-state precesses. Its photoluminescence (PL) emission intensity
as a function of time is given by: 
\begin{equation}
I(t)=I^{0}\exp(-t/\tau_{R}),\label{eq:Radiative decay}
\end{equation}
 where $\tau_{R}$ is its radiative lifetime and $I^{0}$ is the initial
emission intensity.

These three processes of Eqs.\ref{eq:BiE spin}-\ref{eq:Radiative decay}
happen simultaneously, and thus the time dependent circularly polarized
PL intensity of the excited-qubit can be described by: 
\begin{align}
I_{R}(t) & =I^{0}\exp(-t/\tau_{R})\label{eq:PL R detection}\\
 & \cdot\big[1+V^{0}\exp(-t^{2}/T_{2}^{*2})\cos(-\frac{2\pi t}{T_{excited}}+\varphi^{0})\big]/2\nonumber 
\end{align}
 and 
\begin{align}
I_{L}(t) & =I^{0}\exp(-t/\tau_{R})\label{eq:PL L detection}\\
 & \cdot{\normalcolor {\normalcolor }}\big[1+V^{0}\exp(-t^{2}/T_{2}^{*2})\cos(-\frac{2\pi t}{T_{excited}}+\varphi^{0}+\pi)\big]/2\nonumber 
\end{align}
for right and left circularly polarized PL emission, respectively.

The time dependent degree of circular polarization (DCP) is defined
as $DCP(t)=[I_{R}(t)-I_{L}(t)]/[I_{R}(t)+I_{L}(t)]$ , therefore it
is given by:

\begin{equation}
DCP_{p}(t)=V_{P}^{0}\exp(-t^{2}/T_{2}^{*2})\cdot\cos(-\frac{2\pi t}{T_{excited}}+\varphi_{P}^{0})\big]\label{eq:Tomography}
\end{equation}

Since $T_{excited}$ and $T_{2}^{*2}$ can be measured independently,
(see below), one can quite accurately obtain the four variables $V_{H}^{0}$,
$V_{B}^{0},$$\varphi_{H}^{0}$ and $\varphi_{B}^{0}$, by fitting
Eq. \ref{eq:Tomography} to the measured time-dependent {\footnotesize{}$DCP_{p}(t)$}
resulting from the two converting pulses. These four variables accurately
define the initial ground-level-qubit spin state as described by the
three projections $[S_{X}^{0},S_{Y}^{0},S_{Z}^{0}]$ using EQs. \ref{eq:Hconversion}-\ref{eq:Bconversion}.

We prefer to demonstrate the spin state tomography using the DE $\Pi-$system
for two reasons:

(i) The DE and its BIE are an integer spin qubits, and even in the
absence of external magnetic field the exchange interactions between
the carriers remove the degeneracy between their spin up and spin
down states \citep{Bayer2002,Ivchenko2005,Poem2010}. Their eigenstates
are described by $\ket{\pm X_{DE}}=\left(\ket{+Z_{DE}}\pm\ket{-Z_{DE}}\right)/\sqrt{2}$;
and $\ket{\pm X_{BiE}}=\left(\ket{+Z_{BiE}}\pm\ket{-Z_{BiE}}\right)/\sqrt{2}$.
The energy differences between the eigenstates are of an order of
$1\mu eV$ which is smaller than the radiative width of the BIE optical
transition ($\simeq3\mu eV$) and much smaller than the spectral width
of our laser pulse ($\simeq100\mu eV$). Due to these splittings,
the DE and the BIE precession times around the eigenstates' axis $T_{DE}$
and $T_{BiE}$, are about an order of magnitude longer than the radiative
time of the BIE.

(ii) Due to a small mixture between the bright exciton (BE) and dark
exciton (DE) states, it is possible to initialize the DE spin state
with high fidelity using a single optical pulse \citep{Schwartz2015a},
just like the BE \citep{Benny2011}. Initialization of the electron
or the hole spin state is much more complicated \citep{Atature2006,Ramsay_2008}.

Another useful feature of the DE system is that the excitation and
emission of the BIE occur at different wavelengths \citep{Poem2010},
preventing the laser light from blinding the PL detectors.

Fig. \ref{fig:characterization} describes the experiments used to
characterize the BIE as a spin qubit. For the characterization measurements
the DE is prepared in a thermal, totally mixed unpolarized state using
a feeble continuous wave (CW) above-bandgap excitation (457nm) of
the QD \citep{Benny2012,Cogan2018}. A 12-ps-long resonantly tuned
laser pulse then excites the DE to form a BIE. We use three different
polarizations for the pulsed excitation: $\ket R$, $\ket L$, and
$\ket H$. Since the DE-BIE is an optical \textgreek{P}- system, R-
(L-) polarized pulse initializes the BIE in the $\ket{+Z_{BiE}}$
$(\ket{-Z_{BiE}}$) state, while H-polarized pulse results in a totally-mixed
BIE state.

Following its photogeneration, the BIE evolves in time while it recombines
radiatively. Using polarization-sensitive time-resolved PL measurements,
one can readily obtain the BIE precession period ($T_{BIE}$), its
dephasing time ($T_{2}^{*}$), and its radiative decay time ($\tau_{R}$).

In Fig. \ref{fig:characterization}a, we present the measured circularly
polarized PL as a function of time after the pulsed excitation for
three different cases. The blue (red) marks present PL, polarized
Co-(cross-) circular to the excitation pulse, and the green line presents
PL polarized both R and L, following an H-polarized pulse. The solid
black lines present the best-fitted model of Eq. \ref{eq:PL R detection}
and Eq. \ref{eq:PL L detection}. For R- (L-) circularly polarized
conversion pulse the initial BIE phase $\varphi^{0}$ is 0 ($\pi$)
and the visibility $V^{0}$ is 1 (1). For the linear initialization
$V^{0}=0$. In the last case, the PL simply decays radiatively. Therefore,
this measurement is used to determine the radiative lifetime without
any additional fitting parameters.

The BIE precession period and its dephasing are then fitted accurately
using both the time-resolved PL measurements ($I_{R}(t)$ and $I_{L}(t))$
and $DCP$(t) as presented in Fig. \ref{fig:characterization}c. The
measurements are fitted simultaneously and we obtain $T_{BiE}=5.70\pm0.05ns$,
$T_{2}^{*}=5.75\pm0.05ns$ and $\tau_{R}=0.39\pm0.01ns$. As can be
seen in Figs .\ref{fig:characterization}b and d our fitting procedure
agrees with all the measured data points within one standard deviation
of the experimental uncertainty.

We now proceed by first depleting the QD from charges and from the
remaining DE using a 7-ns-long optical pulse as described elsewhere
\citep{Schmidgall2015}. We then write the DE spin state using a 12-ps
optical $\pi-$pulse to an excited DE state \citep{Schwartz2015a}.
About 100-ps later, we convert the DE into the BIE and then apply
time-resolved circular polarization PL measurements to conclude the
state tomography. This cycle is repeated at a 76 MHz rate.

Fig. \ref{fig:Tomographic measurements} displays the results of these
tomographic measurements. In Fig. \ref{fig:Tomographic measurements},
each DE initialization is described by an arrow on the Bloch sphere
on the upper panel. $\ket{\pm X_{DE}}$, $\ket{\pm Y_{DE}}$ and $\ket{\pm Z_{DE}}$
initializations are represented by blue, red and green arrows respectively.
The color-matched solid lines below each initialization represent
the measured DCP as a function of time after the conversion for both
H- and B- conversions. Black solid lines overlaid on the measured
DCP represent the best-fitted model using Eq. \ref{eq:Tomography}.
The differences between the fitted model and the measured DCP, normalized
by the experimental uncertainty, are presented below each curve. The
quality of the fits is evident since every measured point is within
the experimental uncertainty from the calculated one. In the inset
to each figure are schematic descriptions of the BIE spin state after
the conversion pulse and its temporal evolution during its recombination.

One notes that for $\ket{-X_{DE}}$ and $\ket{+X_{DE}}$ initializations,
the DCP for H-conversion is almost flat, which is not surprising since
in these cases the BIE is formed in its $\ket{\pm X_{BiE}}$ eigenstates
that do not evolve in time. Therefore it is nearly impossible to obtain
$\varphi^{0}$ from these measurements. However, since B-polarized
conversion transforms $\ket{\pm X_{DE}}$ to $\ket{\pm Y_{BiE}}$,
which are coherent superpositions of the BiE eigenstates and thereby
precess in time, the DCP is not flat and $\varphi^{0}$ can be easily
obtained. For $\ket{\pm Y_{DE}}$ initializations, B converts $\ket{\pm Y_{DE}}$
to $\ket{\mp X_{BiE}}$, so the DCP signal is flat, while H converts
$\ket{\pm Y_{DE}}$ to $\ket{\pm Y_{BiE}}$, so the DCP is not flat.
For $\ket{\pm Z_{DE}}$ initializations, the DCP for both conversions
is similar, because in both cases $\ket{\pm Z_{DE}}$ is converted
to $\ket{\pm Z_{BiE}}$.

For each DE initialization, we determine the state by fitting the
measured DCP curves of Fig. \ref{fig:Tomographic measurements}, for
both H and B- conversions, to the model of Eq. \ref{eq:Tomography}.
We note that only three fitting parameters $[S_{X}^{0},S_{Y}^{0},S_{Z}^{0}]$
are used to fit 12 time-resolved DCP(t) curves. Fig.\ref{fig:Density matrices}a
shows the results of our fitting procedure, displayed on the DE Bloch
sphere. Color-matched ovals represent one standard deviation of the
experimental uncertainty of the measured-states, while color-matched
arrows represent the six initialized-states. The length of the arrows
represents the degree of polarization of the initializations, independently
measured to be 0.82, due to the finite efficiency of the optical depletion
\citep{Schmidgall2015}.

Finally, we show in Fig. \ref{fig:Density matrices}b the $4\times4$
$\overleftrightarrow{M}$ matrix which maps the initialized DE state,
represented by a $2\times2$ density matrix $\hat{\rho}_{init}$ to
the DE state density matrix that we obtain by state tomography $\hat{\rho}_{tomog}$,
or: 
\[
\hat{\rho}_{tomog}=\overleftrightarrow{M}\cdot\hat{\rho}_{init}.
\]

We obtain the $\overleftrightarrow{M}$ matrix from our measurements
by finding the most probable positive and trace-preserving (physical)
matrix that maps the six initial DE states to the measured ones in
a similar procedure as described in Ref. \citep{Schwartz2016}. The
matrix that we obtain deviates from the expected $4\times4$ identity
matrix $\overleftrightarrow{I}$, also shown in Fig. \ref{fig:Density matrices}b
for comparison. A fidelity of $0.94\pm0.02$ \citep{Jozsa1994} quantifies
the similarity between $\overleftrightarrow{M}$ and $\overleftrightarrow{I}$.
We attribute this deviation to the calibration accuracy of our liquid-crystal
variable-retarder based polarization analyzers \citep{Schwartz2015a}.

In summary, we demonstrate an all-optical measurement method for full
state tomography of electronic spin qubits confined in semiconductor
nanostructures. While previous methods, require the spin to be a part
of an optical $\Lambda-$system, our method is more general since
it relies on an optical $\Pi-$system, typical to long-lived confined
electronic spins, such as conduction band electrons, valence-band
holes, and dark excitons. The ability to perform full state tomography
on electronic spins this way is essential for scaling up hybrid spin-multi
photons graph-states, thereby constituting important step towards
realizations of quantum information-based technologies.

\section*{acknowledgments}

The support of the Israeli Science Foundation (ISF), and that of the
European Research Council (ERC) under the European Union\textquoteright s
Horizon 2020 research and innovation programme (Grant Agreement No.
695188) are gratefully acknowledged.


\input{Complete_state_tomography_of_a_quantum_dot_spin_23.bbl}

\end{document}

%% file: Complete_state_tomography_of_a_quantum_dot_spin_23.bbl
%